\pdfoutput=1
\documentclass[a4paper,UKenglish,cleveref]{lipics-v2021}

\usepackage{mathtools}
\usepackage{xspace}
\usepackage{complexity}
\usepackage{oplotsymbl} 

\AddToHook{env/lemma/begin}{\crefalias{theorem}{lemma}}
\AddToHook{env/corollary/begin}{\crefalias{theorem}{corollary}}
\AddToHook{env/proposition/begin}{\crefalias{theorem}{proposition}}
\AddToHook{env/definition/begin}{\crefalias{theorem}{definition}}
\AddToHook{env/claim/begin}{\crefalias{theorem}{claim}}
\AddToHook{env/remark/begin}{\crefalias{theorem}{remark}}
\AddToHook{env/observation/begin}{\crefalias{theorem}{observation}}

\AtBeginDocument{\hypersetup{pdfsubject={}}}

\newcommand{\ruleset}[1]{\textnormal{\textsc{#1}}\xspace}
\newcommand{\GFjords}{\ruleset{Graph Fjords}}
\newcommand{\TFjords}{\ruleset{Tile Fjords}}
\newcommand{\incidence}{\ruleset{Incidence}}

\newcommand{\tri}{\Delta}
\newcommand{\inftri}{\tri_\infty}
\newcommand{\tridist}{\operatorname{dist}_{\tri}}
\newcommand{\hive}{\hexago}
\newcommand{\infhive}{\hive_\infty}
\newcommand{\set}[1]{\mathcal{#1}}
\newcommand{\rift}[2]{m_{{#1}\leftrightarrow{#2}}}

\title{Fjords is hard on honeycomb trees}

\author{Alfie Davies}
{Department of Computer Science, Durham University, Durham, UK}
{research@alfied.xyz}
{https://orcid.org/0000-0002-4215-7343}
{supported by Leverhulme Trust Research Project Grant RPG-2024-182.}

\author{David C. Kutner}
{Department of Computer Science, Durham University, Durham, UK}
{david.c.kutner@durham.ac.uk}
{https://orcid.org/0000-0003-2979-4513}
{supported by Leverhulme Trust Research Project Grant RPG-2025-400.}

\authorrunning{A.\ Davies and D.\,C.\ Kutner}

\Copyright{Alfie Davies and David C. Kutner}



\category{}
\relatedversion{}

\acknowledgements{
  We would like to thank Henry Austin, Tala Eagling-Vose, and Maximilien
  Gadouleau for insightful discussions during this project.
}

\hideLIPIcs
\nolinenumbers

\begin{document}

\maketitle

\begin{abstract}
  Burke and Hearn showed that a graph generalisation of the expansion phase of
  the board game \emph{Fjords} is \PSPACE-complete on planar graphs, and left
  open whether hardness persists on subgraphs of the hexagonal grid. We resolve
  this question positively: the graph game remains \PSPACE-complete even on
  trees in the honeycomb grid. For the published game with arbitrarily many
  copies of its tiles and pieces, deciding winnability of the expansion phase
  is \PSPACE-complete under the two-player scoring rules of both editions, even
  on positions reachable through legal play in the exploration phase. Our
  instances use only four tile designs besides the three mandated starting
  tiles; both the playable area and its initially unoccupied subgraph are
  subcubic trees, and ties are impossible. The same restrictions yield
  \PSPACE-completeness under normal play, mis\`ere play, and mis\`ere scoring
  (where the lower score wins). Moreover, for every fixed non-negative integer
  $\ell$, hardness persists when the playable area is a uniform
  $\ell$-subdivision of a subcubic tree, giving new forbidden-subgraph
  consequences. Our reduction is from Maker--Breaker \incidence, for which we
  also strengthen the known \PSPACE-completeness result from forests to trees.
\end{abstract}

\medskip
\noindent\textcolor{lipicsGray}{\fontsize{9}{12}\sffamily\bfseries On the use of AI tools:\enskip}{\small
No GenAI tools were used in the preparation of this manuscript, nor in the development of the theory presented herein. 
This disclosure will be updated to describe any use of such tools in the preparation of subsequent versions.
}

\section{Introduction}

Studying the computational complexity of real-world games has an established
tradition in algorithmic game theory: classic two-player games such as
Chess~\cite{fraenkel-lichtenstein:chess}, Go~\cite{lichtenstein-sipser:go},
Hex~\cite{reisch:hex}, and Dots \& Boxes~\cite{buchin.ea:dots-and-boxes};
modern commercial tabletop games such as Hive~\cite{andel-rin:hive}, Battle
Sheep~\cite{burke-ono:battle-sheep}, Push Fight~\cite{bosboom.ea:push-fight},
and single-player UNO~\cite{demaine.ea:uno}; and even video games such as Super
Mario Bros.~\cite{aloupis.ea:nintendo} have all been shown to be
computationally hard under natural generalisations of their rules (see also
Hearn and Demaine~\cite{hearn-demaine:games}).

One subtlety in proving complexity results for such games is the gap between
hardness for arbitrary positions and hardness for \emph{reachable} ones; a
position is reachable if it can be obtained through some sequence of legal
moves from the standard starting position (such as an empty square board in
Go).\footnote{
  Note that we require only the existence of a legal sequence, not a forcing
  one.
}
Proving hardness for reachable positions can add serious difficulty to a
reduction, and is often omitted: for example, Fraenkel and Lichtenstein proved
hardness for arbitrary positions in generalised Chess, writing explicitly that
they ``have not tried to answer questions of reachability'', and that ``it
seems offhand as though players would have a hard time trying to reach [their]
board positions from any reasonable starting
position''~\cite[p.~200]{fraenkel-lichtenstein:chess}. By contrast, Andel and
Rin~\cite[Section~3.10 on p.~3:18]{andel-rin:hive} sketch how their hard Hive
positions can be reached from the initial empty playing field. In this paper,
we establish \PSPACE-completeness for reachable positions of the published
board game \emph{Fjords}.

\emph{Fjords} is a board game designed by Franz-Benno Delonge, first published
in 2005 by Hans im Gl\"uck and Rio Grande Games, and reissued in 2022 by Grail
Games with additional material by Phil Walker-Harding. The rules differ between
the two editions: the reissue extends the game to three and four players and
adds optional rune stones, and even the two-player variant differs from the
original in subtle ways. In particular, in a single round of the 2005 edition,
tied scores result in a drawn round (a full match sums three rounds, with two
round wins breaking a tie), whereas the standard two-player rules of the 2022
edition are played over a single round and award a tied score to the player who
moves second in the final phase.\footnote{
  The 2022 rulebook also includes a ``Classic Variant'' intended to emulate the
  2005 game by using blind tile drawing and summing the scores of three rounds
  without runes. Even this variant differs from the actual 2005 rules: among
  several discrepancies, it does not explicitly overrule the second-player
  tiebreak that the 2022 edition uses in its other variants.
}
Throughout, we follow the original 2005 rules and terminology,\footnote{
  The 2022 reissue changed much of the terminology: `expansion' became
  `settlement', `farms' became `longhouses', `fields' became `vikings', and so
  on.
} noting explicitly where discrepancies between the two editions matter; our
hardness constructions are designed to work under both.

No prior knowledge of \emph{Fjords} is required, and we give all details of the
rules relevant to our proofs below (any omitted details do not affect our
results). Play proceeds in two phases: \emph{exploration} followed by
\emph{expansion}. Both editions begin exploration with the same three fixed
hexagonal starting tiles (shown in
\cref{fig:fjords-tiles-side-by-side,fig:kickstart}). Players then alternate
turns drawing and placing tiles, with placement subject to three geometric
rules that are crucial to our reachability proof:
\begin{enumerate}
  \item
    each tile must be placed adjacent to at least two existing tiles;
  \item
    the shared sides of adjacent tiles must match completely in terrain (the
    three terrain types are \emph{farmland}, \emph{mountain}, and \emph{sea});
  \item
    the land (farmland and mountain) must remain a single connected landmass
    (that is, sea cannot disconnect one landmass from another).
\end{enumerate}
When a player places a tile containing farmland, they may optionally place one
of their \emph{farms} on it. The rules governing how tiles are drawn differ
between the two editions, but this does not affect reachability: in either
edition, there is always some sequence of draws yielding precisely the tile
order required for our construction.

Once no more tiles can be placed, play moves to the expansion phase, with the
opponent of the player who placed the final tile moving first.

In the expansion phase, the tiles and farms remain fixed for the rest of the
game. From this point onwards, there are no chance events and no hidden
information: play has complete and perfect information, making this phase a
combinatorial game. Players alternate claiming unoccupied farmland tiles by
placing one of their \emph{fields} on a tile that is adjacent, via a shared
edge containing farmland, to a tile upon which one of their farms or fields has
already been placed. When a player has no legal moves remaining, they stop
playing, whilst their opponent continues until they too run out of moves.

Once neither player has a legal move remaining, the round ends and each player
scores one point per field placed (farms do not score). In the 2005 edition, a
full match involves three rounds whose scores are summed to determine the
winner; for our reduction, we need only consider determining the winner of a
single round, since we can assume, for example, that the scores are tied
heading into the third and final round.

Burke and Hearn~\cite{burke-hearn:fjords} previously studied a graph
generalisation of the expansion phase, which we call \GFjords. It is played on
a finite, simple graph $G$ with two disjoint subsets of vertices coloured bLue
and Red, respectively (all other vertices are initially uncoloured). Players
Left and Right alternate turns: on Left's turn, she colours an uncoloured
vertex bLue, provided it is adjacent to a bLue vertex; on Right's turn, he
colours an uncoloured vertex Red, provided it is adjacent to a Red vertex. As
is standard in combinatorial game theory, the game ends immediately when the
player to move has no legal move: under the \emph{normal} play convention that
player loses, whereas under the \emph{mis\`ere} play convention that player
wins. The decision problem for \GFjords asks whether Left has a winning
strategy playing second under the given play convention (by symmetry, the
choice of which player moves second is arbitrary).

Burke and Hearn proved that normal play \GFjords is \PSPACE-complete on planar
graphs~\cite[Theorem~3 on p.~407]{burke-hearn:fjords}. Although they did not
say so explicitly, a close inspection of their construction shows hardness on
planar graphs of maximum degree at most~7~\cite[p.~402 and
Figures~15--20]{burke-hearn:fjords}. They asked whether hardness persists on
subgraphs of the hexagonal grid~\cite[Open Problem~2 on
pp.~409--410]{burke-hearn:fjords}, an open question later reiterated by Buchin,
Hagedoorn, Kostitsyna, and van Mulken~\cite[Section~5 on
p.~25:14]{buchin.ea:dots-and-boxes}.

There are two natural grids associated with a hexagonal tiling: the 3-regular
\emph{honeycomb grid}, whose vertices and edges are the corners and sides of
the hexagons, respectively; and the 6-regular \emph{triangular grid}, the
adjacency graph of the hexagonal tiles themselves. Because \emph{Fjords} is
played on hexagonal tiles, the triangular grid is the natural underlying
structure for the board. However, the honeycomb grid is an induced subgraph of
the triangular grid, so proving hardness on subgraphs of the honeycomb grid
establishes hardness for both geometries simultaneously
(\cref{cor:grid-fjords}).

Of course, analysing the expansion phase of the published game directly is
trivially constant-time, given its fixed, finite set of pieces (as is true of
essentially all commercial board games). We therefore generalise \emph{Fjords}
in the natural way, by allowing arbitrarily many copies of the published pieces
(tiles, farms, and fields). Crucially, we do \emph{not} introduce new tile
types: we use the real tile designs, and simply bring a bigger box to fit more
of them in! (Or, alternatively, spend an exorbitant amount on a large number of
standard-sized copies.)

We call this generalisation \TFjords: we are given a board at the start of the
expansion phase, with farms already placed (but no fields yet), enough fields
for each player, and a player to move, and we ask whether the second player can
force a strictly higher score. Our reduction produces positions in which ties
are impossible, so our hardness result holds for winnability under both
published editions. A position is \emph{reachable} if there exists a legal
history of moves in the exploration phase, starting from the three mandated
starting tiles and using a finite multiset of copies of the published tile
designs, that produces the given board, farm placement, and player to move. All
reachability claims hereafter are in this sense just described.

Given a \TFjords position, we define its \emph{playable area} as the graph
whose vertices are farmland tiles, with an edge between two tiles whenever
farmland meets across their shared side (so a player can expand across that
edge), restricted to the connected components containing at least one farm
(tiles outside these components are `inert' and cannot affect play). In
\GFjords, the playable area is simply the union of components containing a
coloured vertex. Deleting the initially coloured vertices (or farm tiles, in
\TFjords) from the playable area yields what we call the \emph{initially
uncoloured} (or, in \TFjords, \emph{initially unoccupied}) \emph{subgraph}.
We can now state our main result:

\begin{theorem}
  \label{thm:reachable-fjords}
  \TFjords is \PSPACE-complete, even on reachable positions (under both
  published editions) in which both the playable area and its initially
  unoccupied subgraph are subcubic trees.
\end{theorem}

In fact, our reduction constructs an explicit legal history of polynomial
length from the three mandated starting tiles, requiring only four additional
tile types (one of which is needed only once). We emphasise that restricting to
four tile designs is a strengthening rather than a shortcut, since any such
instance is also valid when the full published tile set is allowed. Roughly
speaking, two tile types will suffice for us to build the playable area, whilst
the remaining two form a supporting `glove' of mountain tiles from the starting
tiles.

Under the same restrictions as \cref{thm:reachable-fjords}, deciding
winnability is also \PSPACE-complete under both normal and mis\`ere play
(\cref{cor:tile-play}). Moreover, the same construction in \cref{sec:misere}
yields \PSPACE-completeness under \emph{mis\`ere scoring}, in which the
strictly lower score wins; the reader will be very familiar with this objective
if they have ever played against a young child or a very sore loser.

Normal play is the convention most closely aligned with the scoring rules of
the published game, and indeed Burke and Hearn~\cite{burke-hearn:fjords}
considered only normal play. The mis\`ere results are primarily of interest
within combinatorial game theory, where mis\`ere complexity has been much less
studied than its normal play counterpart, despite growing recent interest.

To prove \cref{thm:reachable-fjords}, we first establish hardness for
particular \GFjords instances, and then develop geometric tools for building
boards legally so that our \GFjords construction lifts isomorphically to a
reachable \TFjords position. Our starting point is the following result.

\begin{theorem}
  \label{thm:graph-fjords}
  \GFjords is \PSPACE-complete on subcubic trees whose initially uncoloured
  subgraphs are connected, under both normal and mis\`ere play.
\end{theorem}

We prove the normal play assertion of \cref{thm:graph-fjords} in
\cref{sec:trees} and its mis\`ere counterpart in \cref{sec:misere}, reducing
from Maker--Breaker \incidence, introduced by Bagan, Deschamps, Duch\^ene,
Durain, Effantin, Gledel, Oijid, and
Parreau~\cite[Section~2.1]{bagan.ea:incidence}. In this scoring positional
game, Maker and Breaker alternate claiming vertices of an initially unclaimed
graph until every vertex has been claimed, with Maker scoring one point for
every edge whose endpoints she claims both of; Maker aims to maximise her
score, whilst Breaker aims to minimise it. Given a graph, a starting player,
and an integer `target' $k$, the decision problem asks whether Maker can
guarantee a score of at least $k$. Hilaire, Montfort, and Oijid proved that
Maker--Breaker \incidence is \PSPACE-complete on
forests~\cite[Corollary~24]{hilaire.ea:happy-vertex}.\footnote{
  Hilaire et al.\ use the Breaker-first reduction of Bagan et al.; the
  modification for Maker starting adds an unused quantified variable, which
  preserves acyclicity~\cite[Remark~20]{bagan.ea:incidence}.
}
In \cref{thm:incidence-trees}, we strengthen this result to trees, settling the
case singled out by Bagan et al.~\cite[Section~7]{bagan.ea:incidence}.

The main idea of our reduction in \cref{subsec:gfjords-reduction} is to convert
Maker's \incidence score into moves that Left can save for later in \GFjords:
each vertex Maker claims gives Left a fixed number of saved moves (which we
cancel out using a reserve for Right), and each edge Maker scores saves Left
one additional move. Crucially, the construction leaves us free to choose the
path lengths, yielding an arbitrary fixed uniform $\ell$-subdivision of a
subcubic tree.

In \cref{sec:reachable}, we turn this tree into a reachable position, first by
embedding a uniform subdivision of the tree into the honeycomb grid, and then
by supporting its 5-subdivision with a `glove' of mountain tiles. Since the
initially unoccupied subgraph has odd order, the normal play winner also wins
on score (and ties are impossible).

In \cref{sec:misere}, we obtain our mis\`ere results by adapting the
\emph{poison move} idea of Stewart's reduction for
\ruleset{Hackenbush}~\cite[Theorem~4]{stewart:misere-hackenbush}. Attaching a
single poison vertex with a sufficiently long private path ensures that
claiming the poison gives its owner more private moves than the opponent can
make elsewhere: under mis\`ere play, the opponent therefore runs out of moves
first and wins; under mis\`ere scoring, the player is forced to accumulate a
larger score than their opponent, leaving them in a potentially uncomfortable
social situation.

Finally, in \cref{sec:subdivisions}, we exploit our freedom to vary path
lengths to obtain hardness on uniform $\ell$-subdivisions of subcubic trees for
every fixed $\ell\ge0$. In the forbidden-subgraph framework of Johnson, Martin,
Oostveen, Pandey, Paulusma, Smith, and van
Leeuwen~\cite[Sections~1.1--1.2]{johnson.ea:forbidden-subgraphs}, this places
both \GFjords and reachable \TFjords within the class C23, yielding hardness
for many classes defined by forbidden subgraphs or minors.

\section{Hardness on trees}
\label{sec:trees}

\subsection{Incidence on trees}

We start with the source game for our reduction: Maker--Breaker \incidence.
Maker and Breaker claim vertices in turn, and Maker receives a score of 1 point
for each edge that she claims both endpoints of. As previously discussed,
hardness was already known on forests. For the \GFjords construction it will be
useful to start with a single tree; we could still make the reduction work
directly from forests, but it adds several complications that we can bypass
entirely by briefly proving this theorem first.

\begin{theorem}
  \label{thm:incidence-trees}
  Maker--Breaker \incidence is \PSPACE-complete on trees (regardless of who
  starts), even when the target score is even.
\end{theorem}

\begin{proof}
  We first give a proof for when Breaker starts. Take a forest instance
  $(F, k)$; write $m$ for its number of edges, and $c$ for its number of
  components. Add two new adjacent vertices $u$ and $v$. Add an arbitrary edge
  between $v$ and each component of the forest. Now add $2(m+1)$ private leaves
  to $u$, and $2(2m+c+2)$ private leaves to $v$. Note that the resulting graph
  $T$ is a tree.

  By \cite[Lemma~15 on p.~9]{bagan.ea:incidence}, we may assign half of the
  private leaves added to $u$ to Maker, and the other half to Breaker, without
  changing Maker's optimal score or the player to move (and similarly for the
  private leaves of $v$). Thus, claiming $v$ gives Maker $2m+c+2$ points from
  her private leaves at $v$, and claiming $u$ gives $m+1$ from the private
  leaves there. The remaining unclaimed vertices are $u$, $v$, and the vertices
  of $F$.

  Now consider Breaker's first move. If he claims $v$, then Maker can score at
  most $(m+1)+m$ on what remains. If he claims a vertex distinct from $v$, then
  Maker can claim $v$ and secure $2m+c+2>(m+1)+m$ points. Thus, Breaker will
  start by claiming $v$.

  Now, if Maker responds anywhere distinct from $u$, then Breaker could claim
  $u$, leaving Maker with at most $m$ points. Claiming $u$, however, gives
  Maker $m+1$ points, and hence Maker will respond with $u$. The remaining game
  is precisely the game on $F$, with Breaker to move. Hence, Maker can score at
  least $k$ on $F$ if and only if she can score at least $k+m+1$ on $T$.

  For the proof when Maker starts, the construction is essentially the same,
  except that we add edges from $u$ (rather than $v$) to the components of the
  forest (but the private leaves stay the same), resulting in $T'$. If Maker
  claims $v$ on her first move, then she secures at least $2m+c+2$ points; if
  she plays elsewhere, then Breaker can claim $v$, leaving Maker with at most
  $(m+1)+c+m < 2m+c+2$ points. Thus, Maker will start by claiming $v$. Now, if
  Breaker responds by claiming $u$, then Maker can score at most $m$ further
  points; if he plays elsewhere, then Maker can claim $u$ and secure at least
  $m+1$ further points from $u$'s private leaves. Hence, Breaker will respond
  with $u$, leaving precisely the game on $F$ with Maker to move. It then
  follows that Maker can score at least $k$ on $F$ if and only if she can score
  at least $k+2m+c+2$ on $T'$.

  In either construction, if the resulting target is odd, we may add two more
  private leaves to both $u$ and $v$. By the same argument as above, the
  reward for claiming either vertex increases by exactly one, and so the same
  opening moves remain optimal. Since Maker will claim exactly one of $u$ and
  $v$, her optimal score therefore increases by one, which gives us an
  equivalent instance that has an even target.

  Note that both constructions have linear size and are computable in
  polynomial time. Membership in \PSPACE{} follows trivially, since each move
  claims a vertex.
\end{proof}

\subsection{The reduction to \GFjords}
\label{subsec:gfjords-reduction}

We reduce from Maker--Breaker \incidence on trees, with Breaker starting. Given
a tree $T$ on $n$ vertices and an even target score $k$, we construct a
position of normal play \GFjords in which Right moves first. Left is able to
win from this position if and only if Maker can guarantee at least $k$ points
on $T$. We may assume that $1\le k\le n-1$, since all other targets are
trivial.

For a vertex $v$ of $T$, we introduce an uncoloured \emph{root} vertex $r_v$
and give it two private leaves, one red and one blue. Either player is
therefore able to claim $r_v$, which represents claiming $v$ in the associated
\incidence game.

We then attach a path to $r_v$ with one \emph{attachment vertex} $a_{vu}$ for
each neighbour $u$ of $v$ in $T$. Beyond the final attachment vertex, we
continue the path with a long \emph{padding} path that ends with a blue vertex.

\begin{figure}[htbp]
  \centering
  \includegraphics[width=\linewidth,page=1]{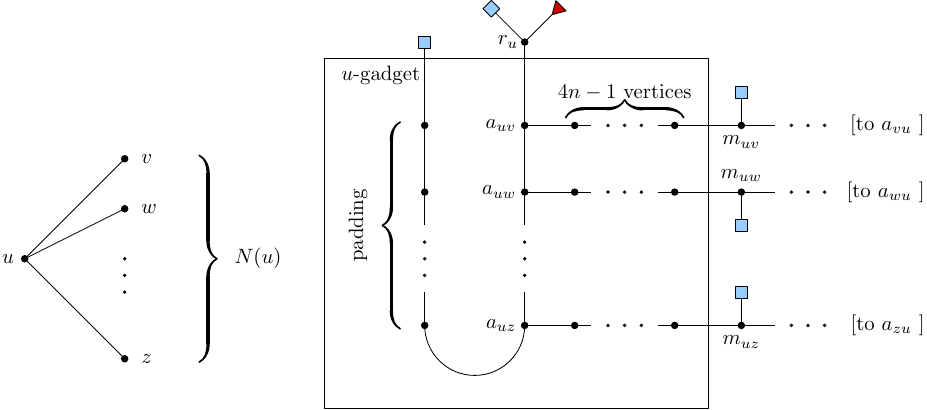}
  \caption{
    Left: a vertex $u$ in an instance of \incidence. Right: the corresponding
    construction for $u$ in \GFjords (before subdivision).
  }
  \label{fig:gfjords-vertex-gadget}
\end{figure}

For each edge $uv$ of $T$, we add an uncoloured \emph{middle} vertex
$m_{uv}=m_{vu}$. We join this middle vertex to both $a_{uv}$ and $a_{vu}$ by
two long uncoloured paths of equal length; we also give $m_{uv}$ a private blue
leaf. Left is therefore able to claim $m_{uv}$ on any turn, provided it remains
uncoloured. If either $r_u$ or $r_v$ is red, then Left will want to claim
$m_{uv}$ to block Right. If she claims both $r_u$ and $r_v$, however, then she
can save this move for later; thus, these saved middle moves will match the
score that Maker can guarantee in \incidence. The construction for a vertex $u$
is illustrated in \cref{fig:gfjords-vertex-gadget}.

To attach Right's reserve, we choose an arbitrary vertex of $T$ and insert one
additional attachment vertex immediately after its last attachment (before its
padding). From this vertex, we add a long path to a new vertex $x$, and from
$x$ another long path to a new vertex $y$. Give $x$ a private blue leaf and $y$
a private red leaf. From $y$, we extend a further path that ends in a single
red vertex. The uncoloured vertices of this last path, excluding $y$ itself,
are \emph{Right's reserve}. Left can claim $x$ to close off the path leading to
$y$, and Right can similarly claim $y$ to keep Left out of his reserve.

This completes our description of the underlying tree, which is our
construction. Its path lengths will be fixed below, using only information from
the \incidence instance. Recall that, for a graph $G$ and an integer $\ell \ge
0$, the \emph{(uniform) $\ell$-subdivision} of $G$, denoted $G^{(\ell)}$, is
obtained by inserting $\ell$ new vertices into every edge of $G$ (i.e.\
replacing each edge by a path of $\ell + 1$ edges). Once those lengths are
fixed, we may choose any $\ell \ge 0$ and take the $\ell$-subdivision of our
entire tree. On the path to each private leaf, we will make all $\ell$ inserted
vertices the same colour as the leaf. Each root will therefore still be able to
be claimed immediately by either player, each middle vertex immediately by
Left, $x$ by Left, and $y$ by Right. On each padding path, we will colour some
final segment blue; similarly, we will colour some final segment of the path
beyond $y$ red to adjust the size of Right's reserve. We leave all other
vertices uncoloured. Since $T$ was a tree and the reserve attaches only once,
the resulting graph is clearly a subcubic tree. Note that removing the coloured
vertices also leaves a (subcubic) tree.

The \emph{gadget} of $v$ consists of the uncoloured vertices on its path of
attachments, its padding, and its paths leading to (but not including) the
middle vertices. It does not include the root; for the single vertex with a
path leading to $x$, its gadget also includes that path (but not $x$ itself).
The number of uncoloured vertices in this gadget is precisely the number of
moves Left saves for herself by claiming $r_v$. By adjusting the coloured
segment at the end of each padding, we make every gadget worth the same amount;
we can then cancel out these saved gadget moves by adjusting Right's reserve.

In particular, letting $c=\lfloor\frac{n}{2}\rfloor$, we know that Maker will
claim precisely $c$ vertices on $T$ in \incidence when Breaker starts (since
the graph starts entirely unclaimed). We will force each gadget in our \GFjords
construction to contain exactly $s$ uncoloured vertices, and Right's reserve to
contain exactly $M=sc+k-1$. Thus, if $L$ is the set of vertices whose roots
Left claims, then her intended supply of moves is
\[
  S(L)=s\cdot|L|+|E(T[L])|.
\]
If she claims Maker's $c$ vertices and scores $k$ edges, then this gives her
precisely one more move than Right has in his reserve. We also arrange for the
total number of uncoloured vertices to be odd, which settles a close case in
Right's strategy and rules out a tied score when it comes to \TFjords later
(which we want to avoid since the published rules deal with ties differently).

\begin{proof}[Proof of \cref{thm:graph-fjords} under normal play]
  A player with a winning \incidence strategy plays by keeping a copy of
  \incidence on $T$: when the strategy calls for playing $v$ on $T$, they claim
  the corresponding root $r_v$. If the opponent claims a root whose vertex is
  still free in the copy, then we record that move; on any other opponent move,
  we assign the opponent an arbitrary unclaimed vertex in the copy. In this
  way, every claimed root has its vertex claimed in the copy (so every reply is
  legal), and whenever the player claims $v$ in the copy, they claim $r_v$.
  Once the copy of \incidence is finished, the player exhausts any remaining
  roots (by claiming them directly until none remain).

  If Left follows Maker's strategy, then $L$ contains Maker's set in the
  \incidence game, and so $S(L)\ge sc+k=M+1$. Similarly, if Right follows
  Breaker's strategy, then $L$ is contained in Maker's set, and so $S(L)\le
  sc+k-1=M$.

  For the remainder of the strategy, we will deal with two cases according to
  who wins the given instance of \incidence. Suppose first that Maker has a
  winning strategy. Left then claims $x$ along with every middle vertex whose
  associated roots are not both blue. Each of these has a blue neighbour by
  construction, and the paths are long enough that she has time to claim them
  (i.e.\ before Right can get there).

  At this point, Right is unable to enter a gadget with a blue root or a middle
  vertex between two such gadgets. Left can therefore save all $S(L)$ of these
  vertices for later.

  Left now plays outside both her supply and Right's reserve for as long as she
  can. Consider a gadget with a red root. Since our graph is a tree and Right
  can enter this gadget only through its root, the path from the root to any
  red vertex in the gadget is entirely red. Now suppose the gadget still
  contains an uncoloured vertex $w$; by construction, $w$ lies on a path from
  the root to a vertex that is already blue (an incident middle vertex, the
  blue end of the padding, or $x$). No vertex from $w$ onwards along this path
  can be red, so the vertex immediately before the first blue vertex beyond $w$
  is uncoloured and gives Left a legal move inside the gadget.

  The same argument applies to the path from $y$ to (the now blue) $x$
  inclusive, since Right can only enter from the $y$ end. Thus, when Left has
  exhausted all moves outside her supply and Right's reserve, Right has at most
  $M$ moves remaining (his reserve). Since Left has at least $M+1$ moves in her
  supply, she wins.

  Now suppose instead that Breaker wins the given instance of \incidence. Right
  next claims $y$ (to secure his reserve), and then colours the attachment path
  of every gadget with a red root. Again, the paths are long enough for Right
  to do this before Left can reach $y$ or any of these attachments. Note that
  Right will not play on his reserve until those are the only moves that
  remain.

  Right has now coloured the entire path from each red root to its last
  attachment. Consider the path from each such last attachment along its
  padding, any path from one of these red attachments to a middle vertex or to
  $x$ (including its far endpoint), or the path from $y$ to $x$ inclusive. Each
  such path starts at a red vertex, and Left can enter it only from its far
  end, so any blue vertices along it are connected to that end. Hence, if the
  path contains an uncoloured vertex, then the first uncoloured vertex from its
  red end is adjacent to a red vertex, giving Right a legal move. Thus, when
  Right has exhausted all of his moves outside his reserve, every remaining
  uncoloured vertex outside it lies in Left's supply, which has size $S(L) \le
  M$.

  When Right first has no move outside of his reserve, the players have made
  equally many moves (since Right started). Each move colours a single vertex,
  and, since the initial number of uncoloured vertices was odd, the number of
  vertices still uncoloured must be odd. Exactly $M$ of them lie inside Right's
  reserve, and at most $M$ lie outside of it. But since the total must be odd,
  we know that strictly fewer than $M$ must lie outside of it, and so Right
  wins.

  It remains only to describe how we adjust the lengths of parts of our
  construction to allow this strategy to work.

  Before subdivision, we make the length of every path from an attachment to a
  middle vertex or to $x$, and also the path from $x$ to $y$, exactly $4n$. We
  then make the length of each padding beyond its last attachment $(4n)^2$, and
  the path beyond $y$ (from which we construct Right's reserve) $(c+1)\cdot
  (4n)^2$. This fixes the tree independently of $\ell$.

  After taking the $\ell$-subdivision, put $s=(\ell+1)(4n)^2$. A gadget with
  $t$ attachments now has $t(\ell+1)$ vertices on its attachment path beyond
  the root, as well as $4n(\ell+1)-1$ internal vertices on each of its $t$
  paths from attachments to middle vertices (and also from an attachment to $x$
  at the one gadget that has this). We can therefore leave
  $s-t(\ell+1)-t(4n(\ell+1)-1)$ padding vertices uncoloured and colour the rest
  blue. Since $1\le t\le n$ and $(4n)^2\ge n(4n+1)+4n$, this leaves at least
  $4n(\ell+1)$ uncoloured vertices and at least one blue vertex.

  Similarly, we leave precisely $M=sc+k-1$ uncoloured vertices beyond $y$ and
  colour the rest red. There is ample room since $k\le n-1<(\ell+1)(4n)^2$.

  Each gadget now has even size $s$, and the reserve has odd size $M$. The
  remaining uncoloured vertices are the $n$ roots, the $n-1$ middle vertices,
  and the $4n(\ell+1)+1$ vertices from $x$ to $y$ inclusive. Their total is
  necessarily even, and so the entire position has an odd number of uncoloured
  vertices.

  Finally, each player following the corresponding \incidence strategy spends
  at most $n$ of their turns playing root vertices. Left then needs at most $n$
  turns to claim $x$ along with the required middle vertices, and Right needs
  at most $1+(\ell+1)(2n-1)$ turns to claim $y$ and every attachment path of a
  gadget with a red root, since there are $2(n-1)+1$ attachment vertices in
  total. In either case, the opponent has necessarily made fewer than
  $4n(\ell+1)$ moves by the end of these preparations. This is too few to cross
  any of the long paths or the uncoloured part of any padding, and so the
  opponent cannot interrupt the strategy we have described.

  The construction takes polynomial time in $n$ and $\ell$. Taking $\ell=0$
  proves hardness on subcubic trees whose initially uncoloured subgraphs are
  connected (and hence subcubic trees themselves). Clearly we are in \PSPACE,
  since each move colours a vertex. Thus, we have the result.
\end{proof}

\section{Reachable \TFjords{} positions}
\label{sec:reachable}

The graph construction specifies which vertices are adjacent and which are
initially coloured. To realise that position in \emph{Fjords}, we must lay
tiles whose farmland gives exactly those adjacencies, and each farm must be
placed by its owner on the turn its tile is laid. We first draw a subdivision
of the tree in the honeycomb grid, and we then give a legal exploration history
that builds the corresponding board and places each farm on its prescribed
tile.

\subsection{Drawing the tree}
\label{subsec:honeycomb-drawing}

Our goal here is to draw some uniform subdivision of a given subcubic tree as a
subgraph of the honeycomb grid. We first draw the tree using odd-length paths
for its edges, and then scale up the grid and adjust the paths so that they all
have the same length. To do this, we require a few definitions.

\begin{definition}[Finite and infinite grids bees feel at home in]
  The infinite triangular grid $\inftri$ is the familiar 6-regular
  2-dimensional lattice. Up to permuting colours, there is a unique proper
  3-colouring $c\colon V(\inftri)\to \{B,G,R\}$. Fixing such a colouring $c$
  and a vertex $o \in V(\inftri)$ with $c(o)=B$ as the origin, we further
  define:
  \begin{itemize}
    \item
      $\tridist(u,v)$ to be the length of a shortest path in $\inftri$ from $u$
      to $v$,
    \item
      the infinite honeycomb grid $\infhive$ to be the subgraph induced in $\inftri$ by
      those vertices assigned $G$ or $R$ under $c$,
    \item
      $\tri_r$ to be the subgraph induced in $\inftri$ by $\{u :
      \tridist(o,u)\le r\}$, and
    \item
      $\hive_r$ to be the subgraph induced in $\infhive$ by $\{u :
      \tridist(o,u)\le r\}$ (using distance in $\inftri$, not $\infhive$).
  \end{itemize}
\end{definition}

\begin{figure}[htbp]
  \centering
  \includegraphics[width=0.7\linewidth,page=2]{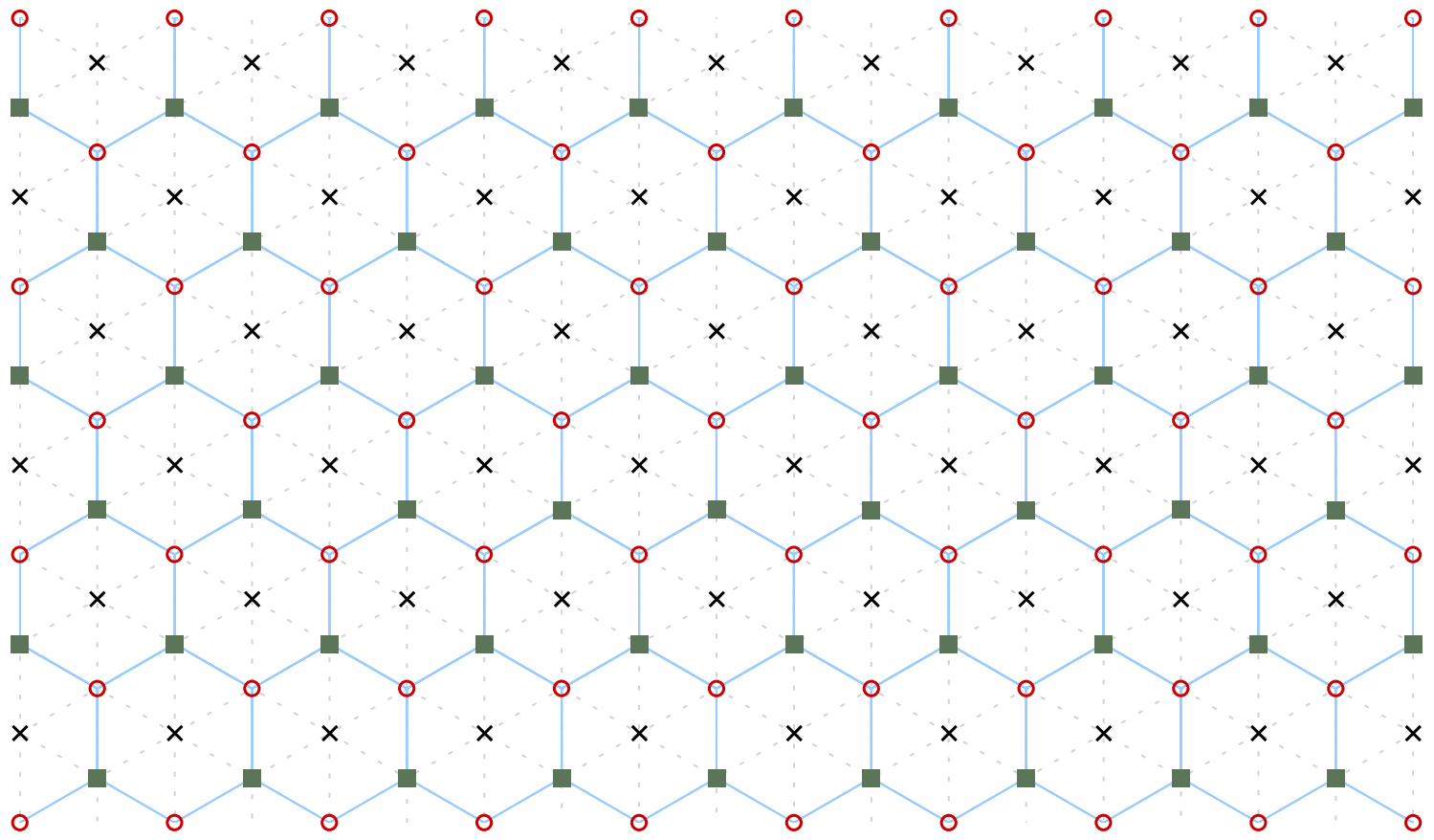}
  \caption{
    (Partial view of) a proper 3-colouring of the infinite triangular grid with
    the colours $\{$black cross, green square, red circle$\}$. The subgraph
    induced by the green squares and red circles (i.e.\ by the removal of all
    black crosses, so also of the dashed grey edges) is the honeycomb grid.
  }
  \label{fig:hive-in-tri}
\end{figure}

\begin{figure}[htbp]
  \centering
  \includegraphics[width=0.7\linewidth,page=3]{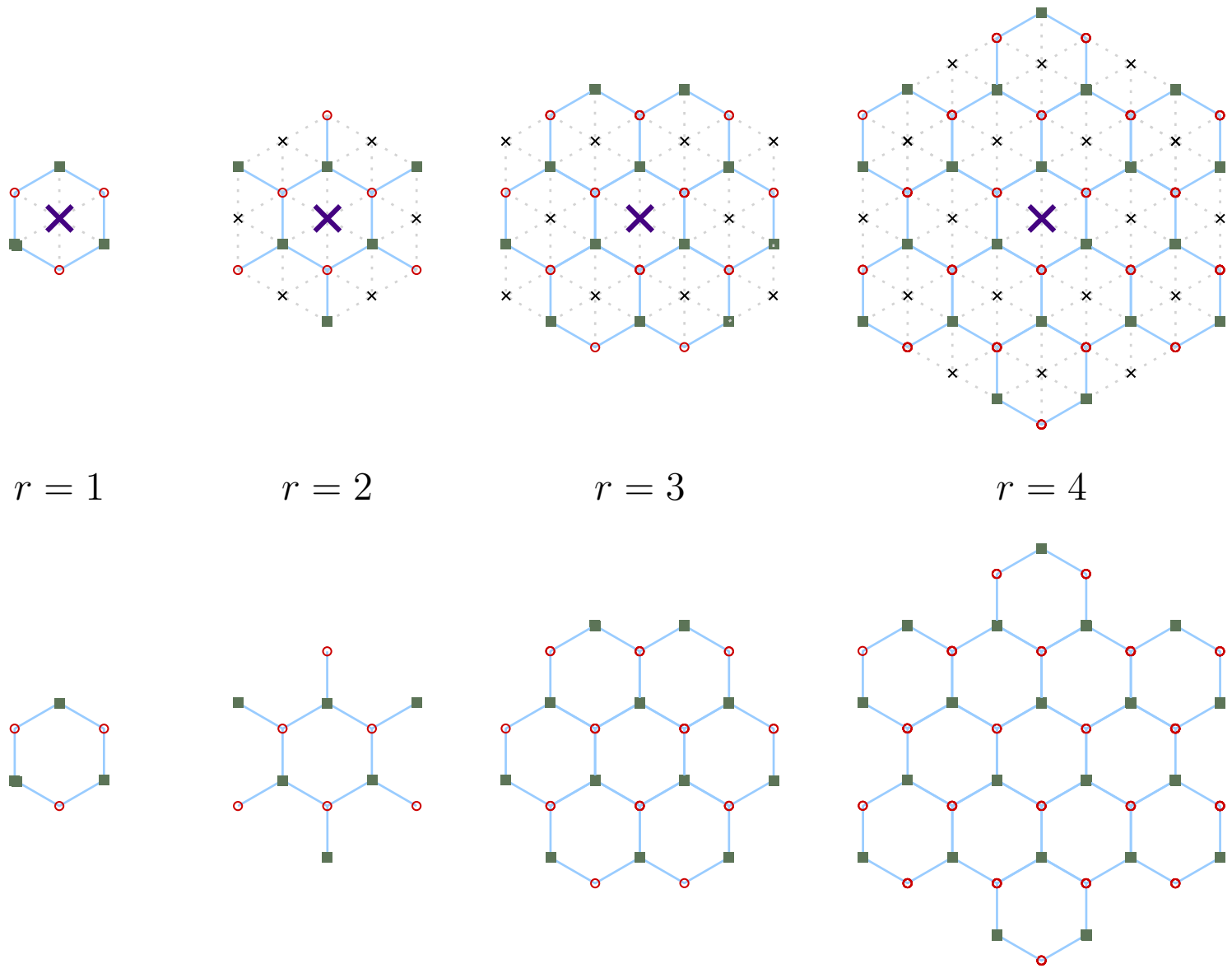}
  \caption{
    Top and bottom, left to right: $\tri_r$ and $\hive_r$, respectively, for
    $r=1,2,3,4$. Each $\hive_r$ is induced in $\tri_r$ by removing all the
    vertices marked with a cross (including the origin, marked with a large
    blue cross), and thus all the dashed grey edges at the top of the figure.
  }
  \label{fig:small-tri-and-hive}
\end{figure}

The grids $\inftri$ and $\infhive$ are illustrated in \cref{fig:hive-in-tri},
and small examples of $\tri_r$ and $\hive_r$ are shown in
\cref{fig:small-tri-and-hive}.

\begin{definition}[Syrupy subgraph]
  A graph $G$ is a \emph{syrupy subgraph} of $H$ if $H$ contains, as a
  subgraph, a graph obtained from $G$ by replacing each edge with a path of odd
  length. Such an embedding $m\colon G \to H$ is called a \emph{syrupy subgraph
  embedding}; we write $m(v)$ for the vertex of $H$ representing $v \in V(G)$
  and $m(e)$ for the path in $H$ representing $e \in E(G)$.
\end{definition}

The reader may find it useful to their intuition to note that if $G$ is a
syrupy subgraph of $H$, then $G$ is a topological minor of $H$, though the
converse may not hold. (This observation is not necessary for our proofs.)

\begin{definition}[Rifting]
  Let $m\colon G \to \hive_r$ be a syrupy subgraph embedding, and let $u$ and
  $v$ be adjacent vertices of $\infhive$. We define the\footnote{
    Strictly, this should be `a' rifting, since an arbitrary choice of $w$
    allows two different riftings at $uv$.
  } \emph{rifting} of $m$ at $uv$, denoted by $\rift u v$, as follows. Let
  $\vec{a} = v - u$ and $\vec{b} = w - u$ for an arbitrary neighbour $w \ne v$
  of $u$ in $\infhive$; then every vertex of $\infhive$ has unique integer
  coordinates $x\vec{a} + y\vec{b}$ relative to $u$. Let the set $N$ of
  \emph{Northern} vertices be those whose coordinates satisfy $x > y$ (and call
  the remaining vertices \emph{Southern}), and define $\phi\colon V(\infhive)
  \to V(\infhive)$ by
  \[
    \phi(z) =
    \begin{cases*}
      z + \vec{a} - \vec{b} & if $z \in N$, \\
      z & otherwise.
    \end{cases*}
  \]
  We set $\rift u v(s) = \phi(m(s))$ for each $s \in V(G)$. For each edge
  $e \in E(G)$, we obtain the path $\rift u v(e)$ by applying $\phi$ to every
  vertex of $m(e)$ and replacing each edge $zz'$ along $m(e)$ whose images
  $\phi(z)$ and $\phi(z')$ are no longer adjacent by the unique shortest path
  of length $3$ between them across the two newly inserted rows.
\end{definition}

As illustrated in \cref{fig:hive-game,fig:rifting}, distinct edges between the
Northern and Southern parts are replaced by internally disjoint $3$-edge paths
using only the newly introduced vertices, and every vertex moves by at most $2$
in $\inftri$. Thus, each path in the image of $\rift u v$ remains simple and of
odd length, so $\rift u v\colon G \to \hive_{r+2}$ is again a syrupy subgraph
embedding. Moreover (and critically for our proof), $\phi(u) = u$, and if the
edge $uv$ is not used by any path in the image of $m$, then $v$ does not appear
in the image of $\rift u v$.

\begin{figure}[htbp]
  \centering
  \includegraphics[width=\linewidth,page=4]{figures.pdf}
  \caption[Rifting, illustrated with tiles from the board game Hive.]{%
    Rifting, illustrated with tiles from the board game \emph{Hive}. Photo
    credit: Teo Kutner.
    \medskip
    \begin{itemize}
      \item[(a)]
        View of $\infhive$ (partial, due to budget constraints). White tiles
        represent vertices, black tiles represent `non-vertices' which are
        removed from $\inftri$ to induce $\infhive$.
      \item[(b)]
        Adjacent vertices $u$ (ladybug) and $v$ (beetle) in $\infhive$.
      \item[(c)]
        Partition (akin to tectonic rifting) of $\infhive$ into Northern and
        Southern parts.
      \item[(d)]
        Further drift.
      \item[(e)]
        Introduction of two new rows of vertices between the two parts.
      \item[(f)]
        Gluing of the Northern and Southern parts along these new vertices.
      \item[(g)]
        The new (ladybug)--(ant)--(bee)--(beetle) shortest path between $u$
        (ladybug) and $v$ (beetle).
    \end{itemize}%
  }
  \label{fig:hive-game}
\end{figure}

\begin{figure}[htbp]
  \centering
  \includegraphics[width=0.8\linewidth,page=5]{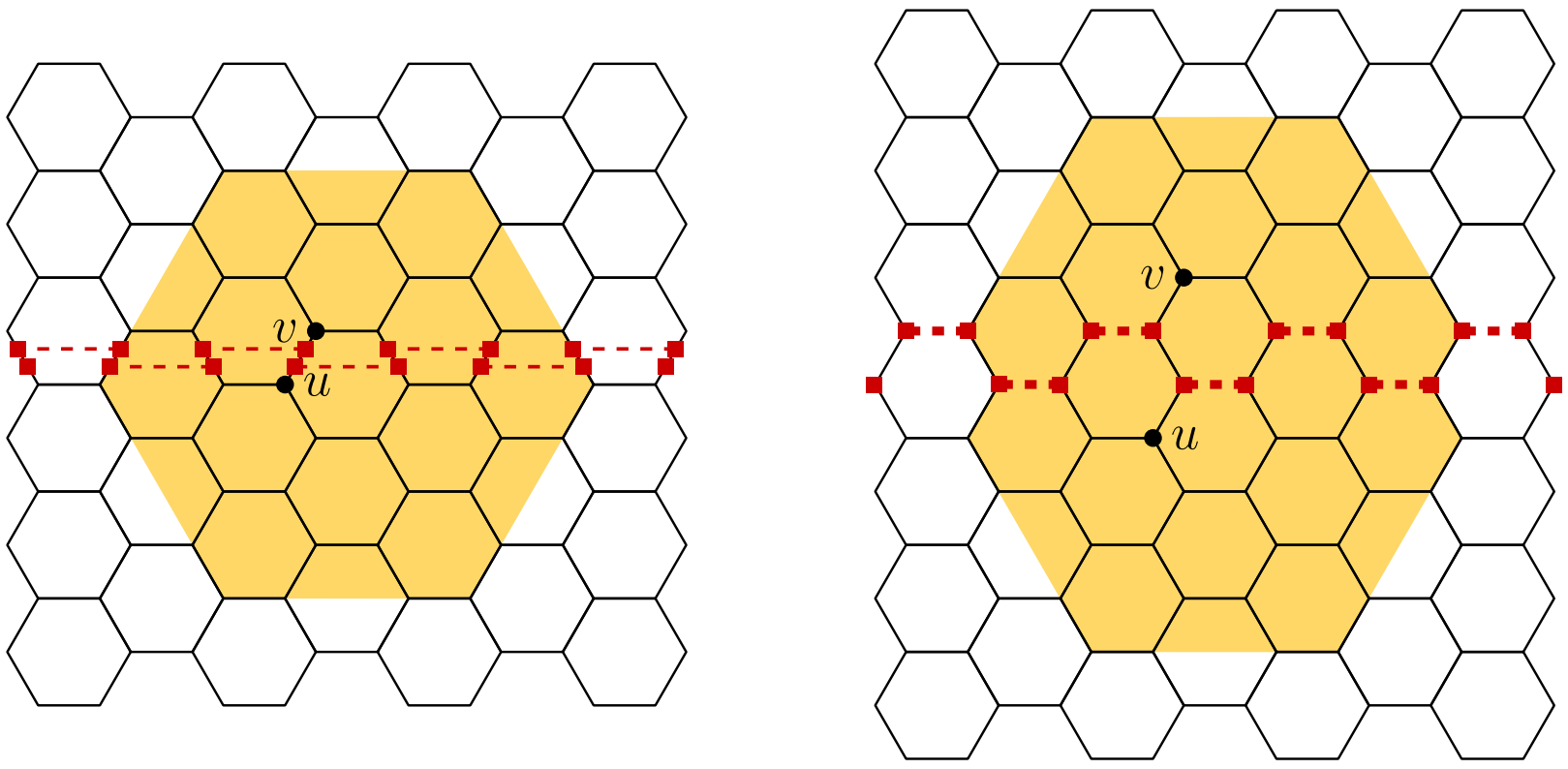}
  \caption{
    Rifting increases the radius of the smallest $\hive_r$ containing the
    entire image of the embedding by at most $2$. Note that if all vertices and
    paths in the image of $m$ are in the shaded yellow area on the left
    pre-rift, then all vertices and paths in the image of $\rift u v$ are in
    the shaded yellow area on the right post-rift.
  }
  \label{fig:rifting}
\end{figure}

\begin{lemma}
  \label{lem:honeycomb-drawing}
  If $T$ is a subcubic tree on $n$ vertices, then for some even $r = O(n^2)$
  with $r \equiv 2 \pmod{3}$, the subdivision $T^{(r^2)}$ is a subgraph of
  $\hive_{(4n+1)r}$, and such an embedding can be constructed in polynomial
  time.
\end{lemma}
\begin{proof}
  Our proof is in two movements. First, we show that $T$ is a syrupy subgraph
  of $\hive_{2n}$, and from there we prove the main statement of the lemma.

  \begin{claim}
    If $T$ is a subcubic tree on $n$ vertices, then $T$ is a syrupy subgraph of
    $\hive_{2n}$.
  \end{claim}
  \begin{claimproof}
    We proceed by induction on $n$. The claim is trivial for $n = 1$.

    Suppose the claim holds for all subcubic trees on $n-1$ vertices, and let
    $T'$ be a subcubic tree on $n \ge 2$ vertices. Let $l \in V(T')$ be a
    leaf of $T'$, let $p$ be its unique neighbour, and let $T = T' - l$. By
    induction, there is a syrupy subgraph embedding
    $m\colon T \to \hive_{2(n-1)}$.

    Write $u = m(p)$. Since $p$ has at most two neighbours in $T$ and no other
    path in the image of $m$ contains $u$, at most two of the three edges incident to
    $u$ in $\infhive$ belong to paths in the image of $m$. Choose a neighbour
    $y$ of $u$ in $\infhive$ such that the edge $uy$ is not used by any path in
    the image of $m$, and take the rifting
    $\rift u y\colon T \to \hive_{2n}$. Then $y$ does not appear in the image
    of $\rift u y$, and
    $\tridist(o, y) \le \tridist(o, u) + 1 \le 2(n-1) + 1 < 2n$, so
    $y \in V(\hive_{2n})$.

    Extending $\rift u y$ by setting $m'(l) = y$ and $m'(pl)$ to be the single
    edge $uy$ gives a syrupy subgraph embedding $m'\colon T' \to \hive_{2n}$,
    completing the induction.
  \end{claimproof}

  Before we can proceed, we need some properties of $\hive_r$ when $r \equiv 2
  \pmod{3}$ is even and $r \ge 8$. The graph $\hive_r$ has exactly six vertices
  of degree one, at its corners. We call two opposite corners $u$ and $v$
  \emph{diametric}; their distance in $\hive_r$ is $d = \frac{8}{3}(r-2) + 5$,
  which is odd and linear in $r$. On the other hand, as illustrated in
  \cref{fig:diametric-paths}, there also exist simple $u$--$v$ paths of length
  at least $r^2 + 1$ whose internal vertices all lie in $\hive_{r-1}$. By
  adjusting the vertical segments (with a small local detour when needed), it
  is straightforward to construct in polynomial time such a path of any odd
  length $L$ with $d \le L \le r^2 + 1$ (even lengths are impossible since
  $\hive_r$ is bipartite). Keeping the internal vertices in $\hive_{r-1}$ lets
  us fit copies of $\hive_r$ together later without overlaps; we call all such
  paths \emph{diametric}.

  \begin{figure}[htbp]
    \centering
    \includegraphics[width=0.8\linewidth,page=6]{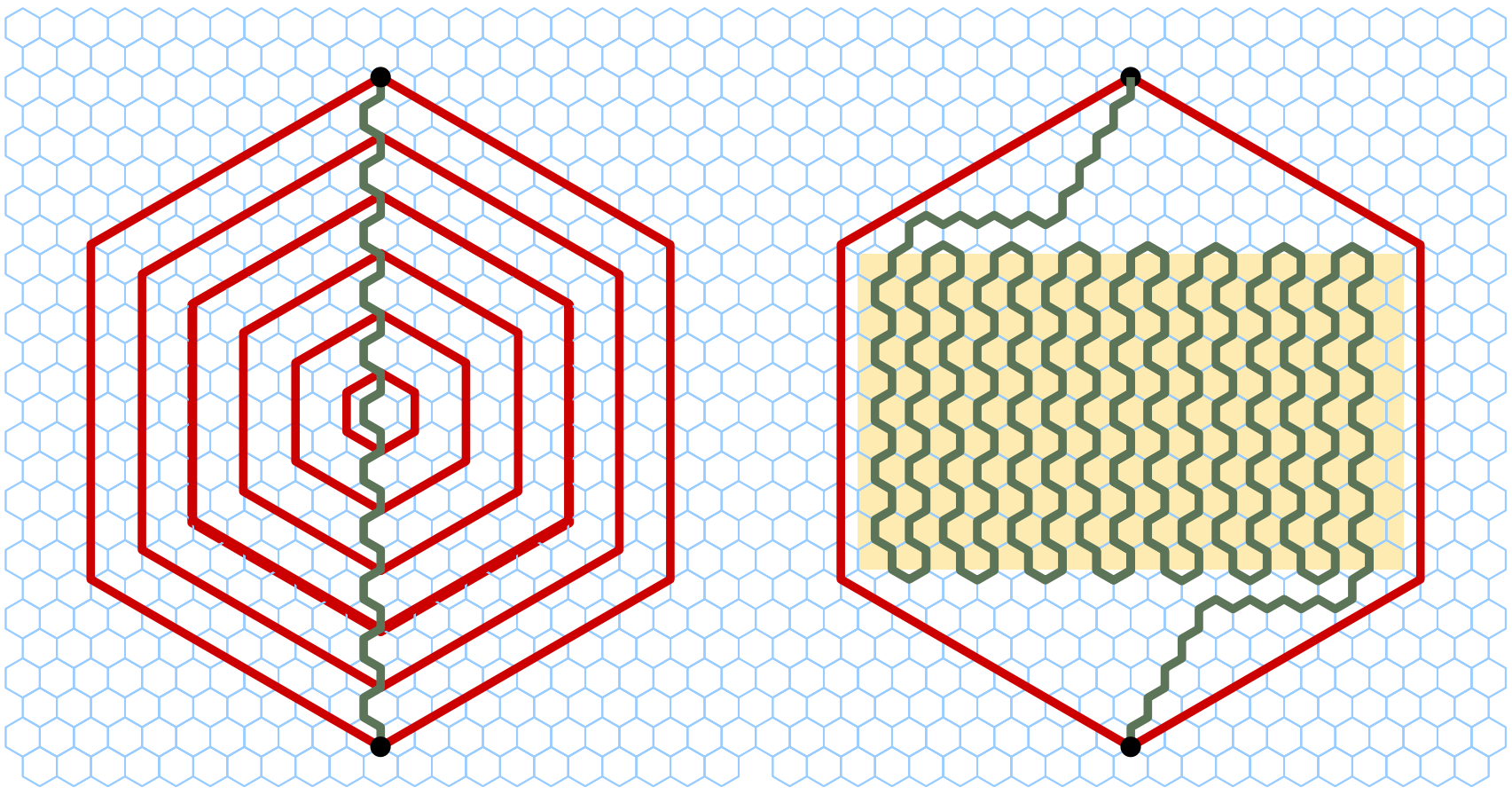}
    \caption{
      Left: the shortest path between diametric vertices in $\hive_r$ with
      small $r \equiv 2 \pmod{3}$. Right: a long path between diametric
      vertices in $\hive_r$ winding through the shaded yellow box, of length at
      least $r^2 + 1$.
    }
    \label{fig:diametric-paths}
  \end{figure}

  \begin{claim}
    If $T$ is a subcubic tree on $n$ vertices, $m\colon T \to \hive_{2n}$ is a
    syrupy subgraph embedding of $T$, and $r \equiv 2 \pmod{3}$ is even with
    $r \ge 3|V(\hive_{2n})|$, then $T^{(r^2)}$ is a subgraph of
    $\hive_{(4n+1)r}$.
  \end{claim}

  \begin{claimproof}
    If $n = 1$, then the claim is trivial; assume $n \ge 2$, so that $r \ge
    3|V(\hive_{2n})| \ge 8$. Every path in the image of $m$ is a simple path in
    $\hive_{2n}$, so its odd length $k$ satisfies $1 \le k < |V(\hive_{2n})|$;
    since $d < 3r$, we obtain $kd < 3r|V(\hive_{2n})| \le r^2$.

    As illustrated in \cref{fig:squigs}, we scale up $\hive_{2n}$ by a factor
    of $2r$ and replace each edge used by a path in the image of $m$ with a
    diametric path in a copy of $\hive_r$ joining its scaled endpoints. Since
    distinct copies of $\hive_r$ have disjoint interiors $\hive_{r-1}$, these
    diametric paths meet only at those endpoints, and the whole construction
    lies inside $\hive_{(4n+1)r}$.

    For each path $m(e)$ of odd length $k$, we replace one edge by a diametric
    path of length $L = (r^2 + 1) - (k - 1)d$, and each of the remaining $k-1$
    edges by a diametric path of length $d$. Since $k$ is odd and $r$ is even,
    $L$ is odd, and $kd < r^2$ implies $d \le L \le r^2 + 1$. Each edge of $T$
    is therefore represented by a path of length $(k-1)d + L = r^2 + 1$,
    yielding $T^{(r^2)}$ as a subgraph of $\hive_{(4n+1)r}$.
  \end{claimproof}

  \begin{figure}[htbp]
    \centering
    \includegraphics[width=\linewidth,page=7]{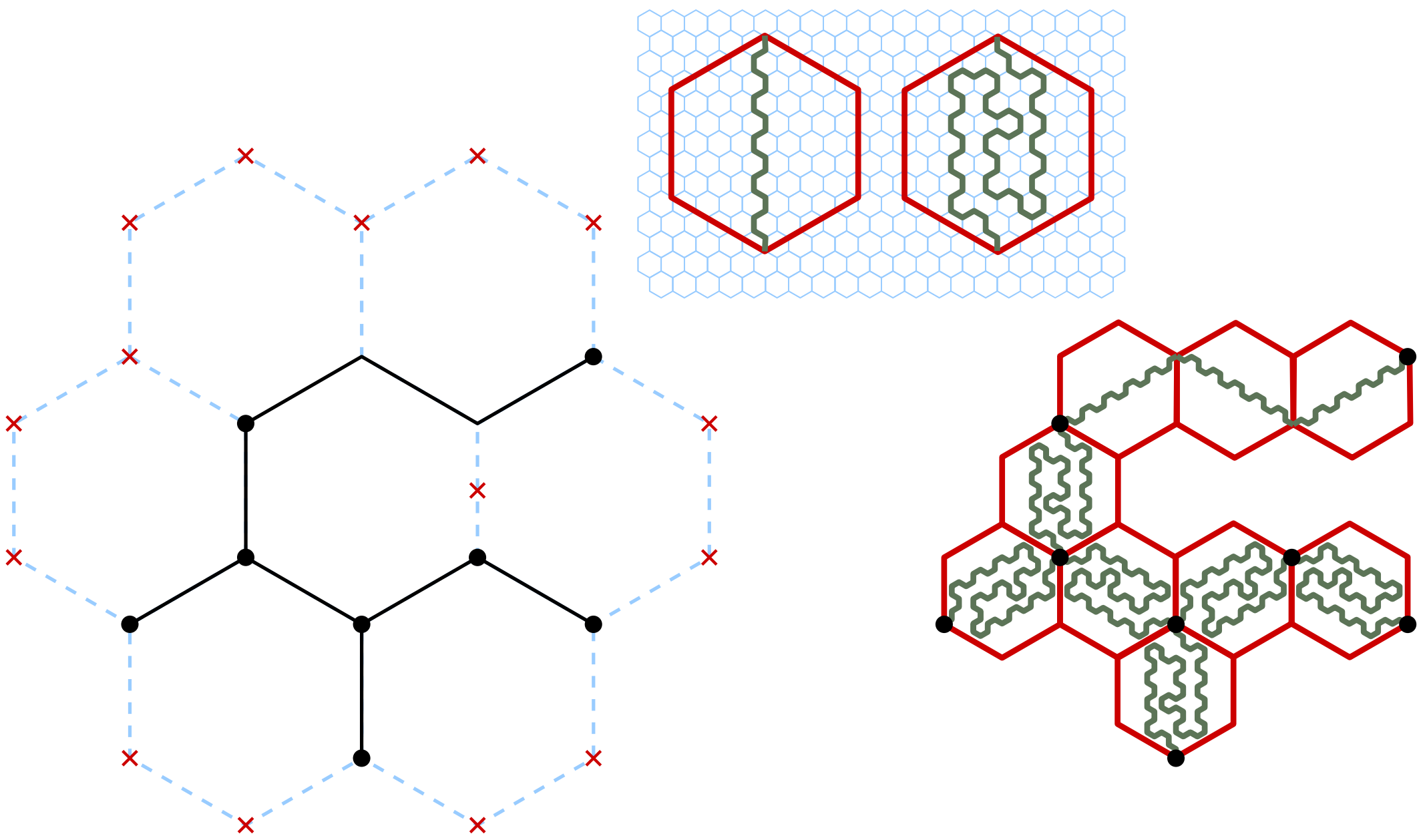}
    \caption{
      Left: a tree $T$ (black, solid) which is a syrupy subgraph of $\hive_3$
      (blue, dashed). Top: diametric paths in $\hive_8$ of length $21$ and
      $63$. Right: using common path length $63$ gives $T^{(62)}$ as a subgraph
      of $\hive_{2 \cdot 8 \cdot 3 + 8}=\hive_{56}$.
    }
    \label{fig:squigs}
  \end{figure}

  Choosing the smallest even $r \equiv 2 \pmod{3}$ with
  $r \ge 3|V(\hive_{2n})|$ gives $r = O(n^2)$ (since $|V(\hive_{2n})| = O(n^2)$).
  Observe that the entire proof is constructive: rifting to build the syrupy
  subgraph embedding, computing $r$ and $d$, and constructing the diametric
  paths in $\hive_{(4n+1)r}$ can all be performed in polynomial time.
\end{proof}

\subsection{Building the board}

At a leaf, only one neighbouring tile belongs to the playable tree, and yet the
rules dictate that laying a tile requires two adjacent tiles to already be
placed. How do we overcome this? The trick we employ is to use a `glove' of
mountain tiles that act as supports, providing any additional required
neighbours. The mountain tiles are inert: they do not
affect either player's available moves during the expansion phase.

\begin{figure}[htbp]
  \centering
  \includegraphics[width=\linewidth,page=8]{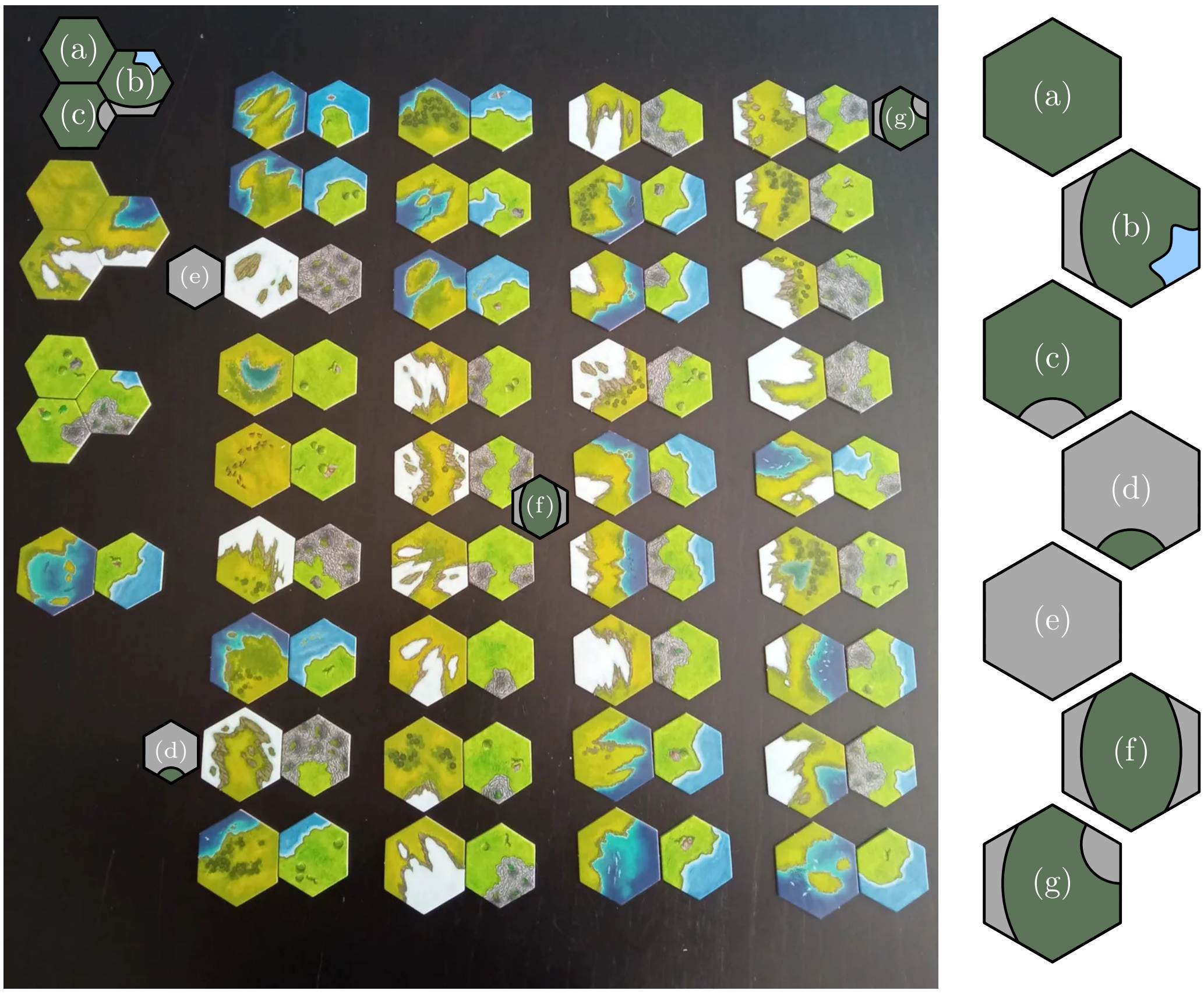}
  \caption{
    The \emph{Fjords} tiles (2005 and 2022 editions) alongside our diagrammatic
    notation: the starting tiles~(a)--(c), the kickstart tile~(d), the mountain
    tile~(e), the straight tile~(f), and the fork tile~(g). Image credit: Roi
    Espino, \url{https://boardgamegeek.com/image/6859167/fjords}.
  }
  \label{fig:fjords-tiles-side-by-side}
\end{figure}

\begin{figure}[htbp]
  \centering
  \includegraphics[width=\linewidth,page=9]{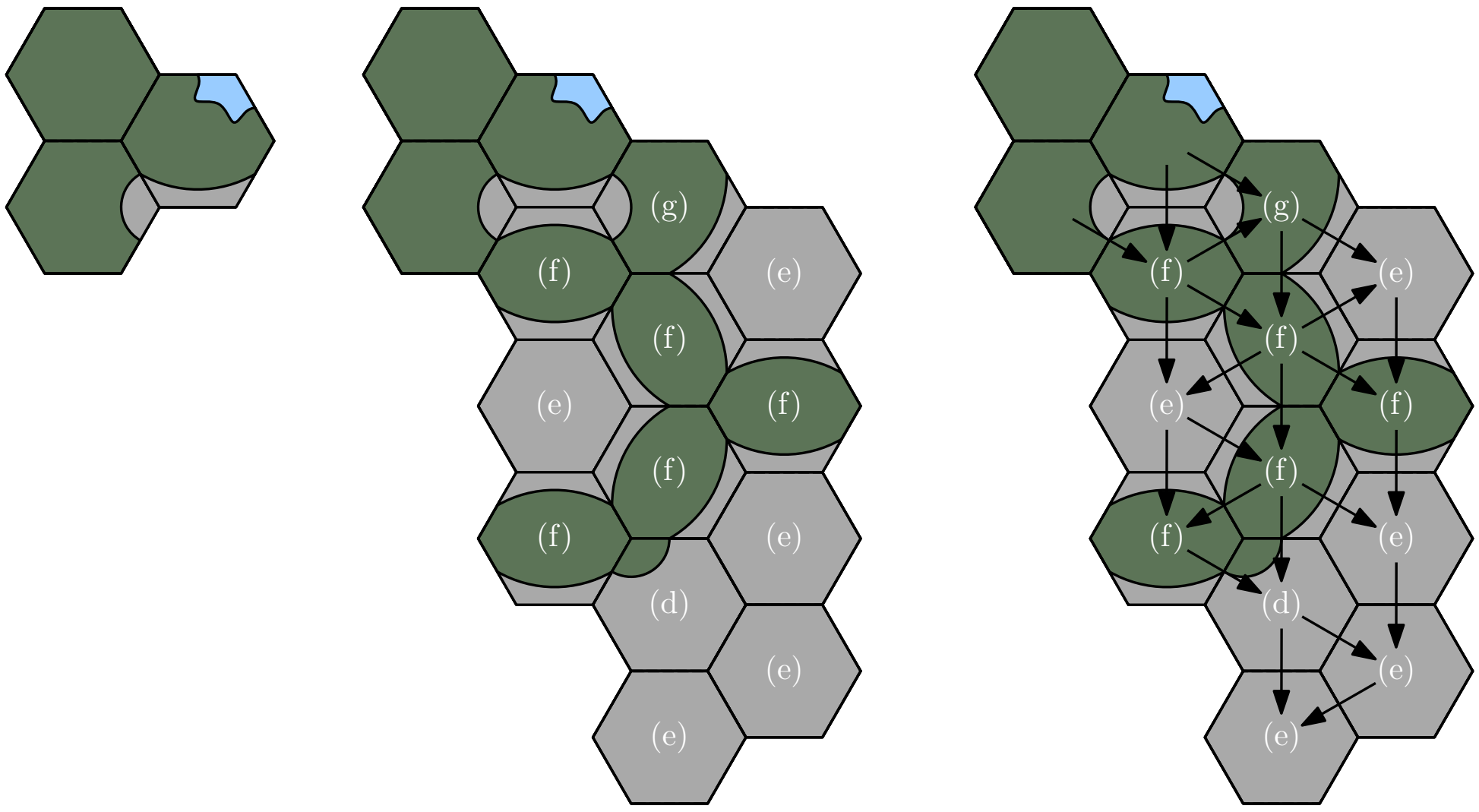}
  \caption{
    Left to right: the starting configuration, our `kickstart' configuration,
    and an acyclic digraph whose arrows give at least two earlier neighbours
    for each placed tile.
  }
  \label{fig:kickstart}
\end{figure}

The construction uses only four tile designs besides the three mandated
starting tiles, as shown in \cref{fig:fjords-tiles-side-by-side}, and it also
allows us to prescribe who moves first in the expansion phase. Starting from
the three mandated tiles, the `kickstart' configuration in \cref{fig:kickstart}
legally produces two adjacent mountain tiles, from which we can grow the glove.

\begin{lemma}[Glove lemma]
  \label{lem:glove}
  If $T$ is a finite subtree of the honeycomb grid $\infhive$ with at least one
  edge, then its 5-subdivision $T^{(5)}$ with any non-empty placement of farms,
  and either choice of player to move first in the expansion phase, occurs as
  the playable area of a reachable \TFjords position (under both published
  editions). Moreover, a legal exploration history of polynomial length using
  only the four tile designs~(d)--(g) besides the three starting tiles can be
  constructed in polynomial time.
\end{lemma}

\begin{proof}
  As illustrated in \cref{fig:board-building-subgraph}, the infinite honeycomb
  grid $\infhive$ can be realised using fork tiles~(g) at its vertices and five
  alternating straight tiles~(f) along each edge, with matching terrain at
  every join and no unintended adjacencies. Deleting the tiles outside $T$
  leaves a placement of tiles whose farmland graph is precisely $T^{(5)}$.

  \begin{figure}[htbp]
    \centering
    \includegraphics[width=\linewidth,page=10]{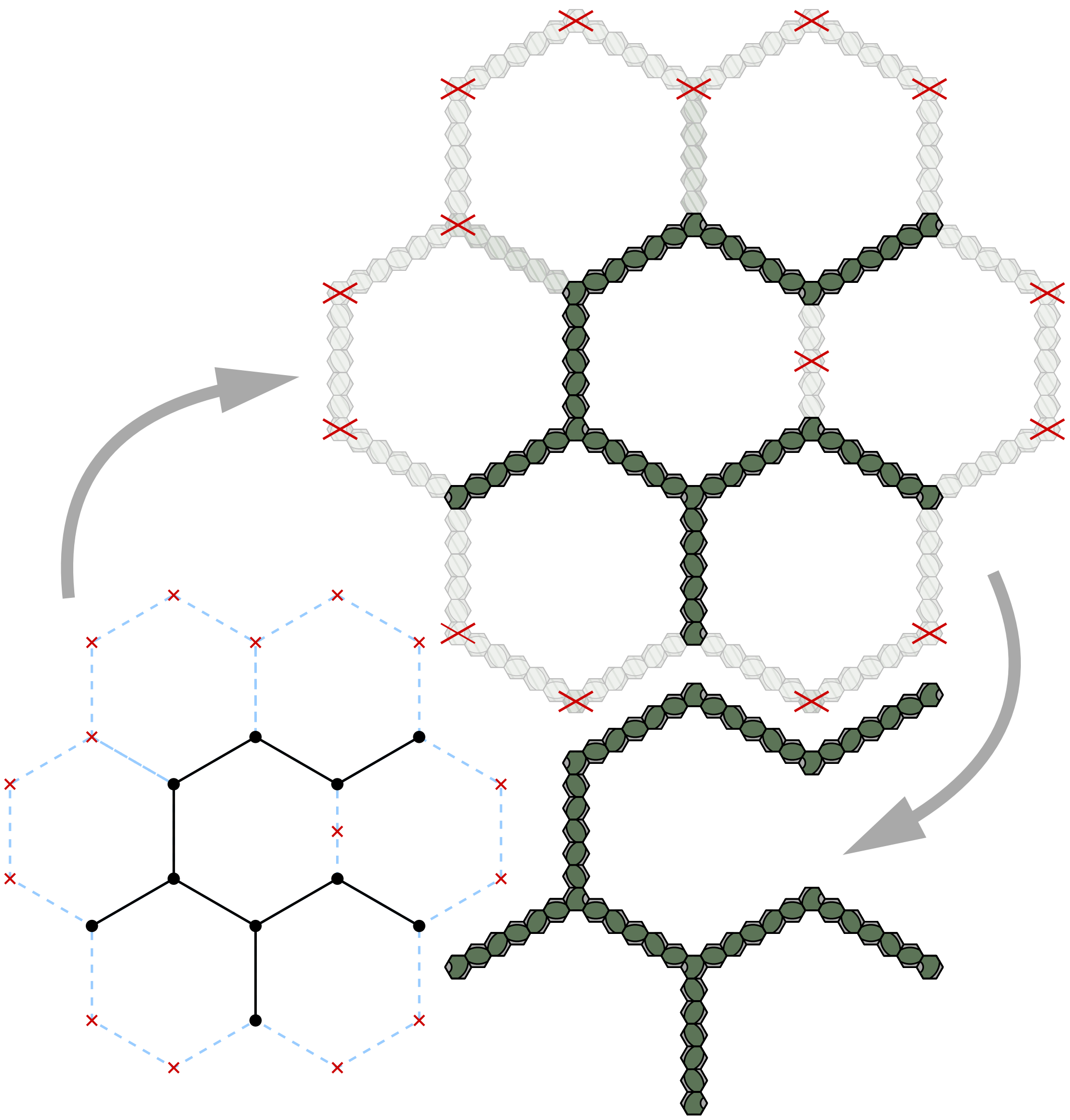}
    \caption{
      Clockwise from bottom left: a tree $T \subseteq \hive_3$, the
      corresponding tile realisation of $\hive_3$ with unused parts marked for
      deletion, and the resulting playable area $T^{(5)}$.
    }
    \label{fig:board-building-subgraph}
  \end{figure}

  We now build the board in four phases, as shown in
  \cref{fig:board-building-glove}. Call a cell outside $T^{(5)}$
  \emph{forbidden} if it touches farmland on $T^{(5)}$ (shaded yellow in
  \cref{fig:board-building-glove}); leaving all forbidden cells empty ensures
  that every later contact with a mountain tile matches terrain. Surround
  $T^{(5)}$ by a hexagonal disc $D$ with a sufficiently wide clear margin
  (since $T$ is connected, $D$ has $O(|V(T)|^2)$ cells), and place the
  kickstart (\cref{fig:kickstart}) outside $D$ so that its final pair of
  adjacent mountain tiles~(e) feeds into the boundary of $D$, with the rest of
  the kickstart behind them.
  \begin{itemize}
    \item
      \emph{Phase~1 (fitting the glove):} build the kickstart, and then
      greedily place a mountain tile~(e) in every non-forbidden cell of
      $D \setminus T^{(5)}$ touching at least two earlier mountains.
    \item
      \emph{Phase~2:} place the three middle straight tiles~(f) of each edge of
      $T$ (those not adjacent to a fork); each touches two mountain supports
      from Phase~1.
    \item
      \emph{Phase~3:} place the remaining two straight tiles~(f) of each edge;
      each touches a mountain support from Phase~1 and its adjacent straight
      tile from Phase~2.
    \item
      \emph{Phase~4:} place every fork tile~(g); each internal fork touches two
      or three straight tiles from Phase~3, whilst each leaf touches its unique
      straight neighbour and a mountain support from Phase~1.
  \end{itemize}

  \begin{figure}[htbp]
    \centering
    \includegraphics[width=\linewidth,page=11]{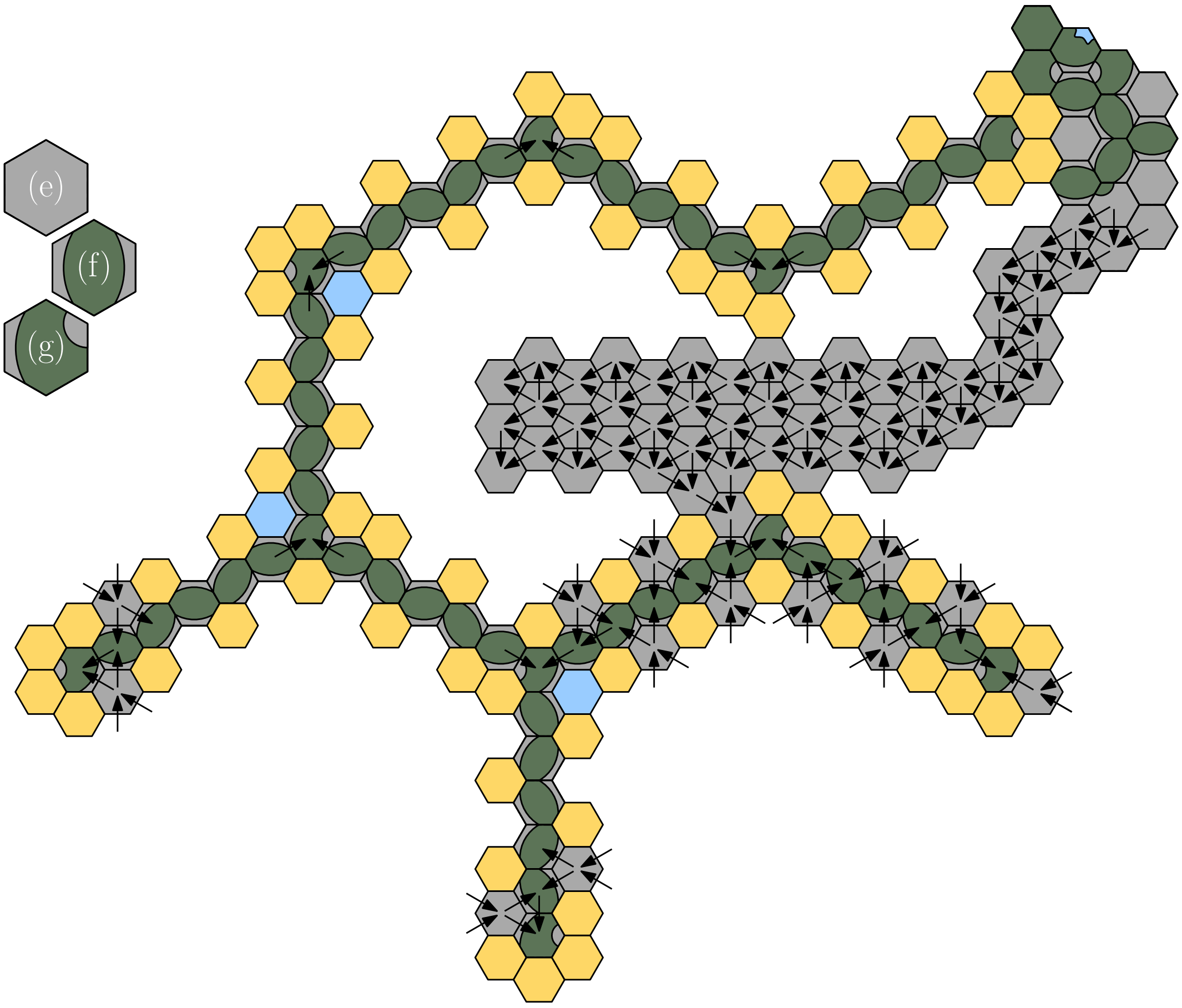}
    \caption{
      Fitting the glove around $T^{(5)}$ from the kickstart. Cells adjacent to
      farmland on $T^{(5)}$ (yellow) must remain empty; inner corners (light
      blue) are not needed.
    }
    \label{fig:board-building-glove}
  \end{figure}

  To see that Phase~1 supplies all of the mountain supports used above, observe
  that two adjacent mountains can grow along any strip at least two tiles wide
  (each new tile touching two earlier ones). Starting from the kickstart, the
  glove fills the clear margin around $T^{(5)}$ and enters every honeycomb face
  meeting $T$: since $T$ is a tree, no cycle can enclose a face, and every
  unused honeycomb edge leaves a gap at least two tiles wide for the mountains
  to pass through. As shown in \cref{fig:board-building-glove}, once these face
  interiors are filled, the glove reaches the four shared supports along every
  edge of $T$ as well as the support on the mountain side of every leaf~(g) (at
  a leaf, two of its three incident honeycomb edges are unused, so its mountain
  support is never trapped in an inner corner like the light-blue cells of
  \cref{fig:board-building-glove}).

  To assign farms to their prescribed owners and choose who moves first in
  expansion, we reserve a two-wide strip of mountain tiles~(e) growing outwards
  from the side of $D$ opposite the kickstart. Since a two-wide strip can be
  extended one tile at a time (each new tile touching two earlier mountains),
  we may append a single mountain tile whenever we need to switch the player to
  move. Placing no farms on the kickstart, whenever a tile of $T^{(5)}$
  requires a farm of player~$P$ and $P$ is not to move, we first append one
  mountain to this strip so that $P$ lays the tile and their farm. Once
  $T^{(5)}$ is complete, we append one final mountain if needed so that the
  opponent of the prescribed first player places the last tile.

  None of the four designs~(d)--(g) contains sea and every placement touches at
  least two earlier tiles, so adding them never disconnects the landmass. The
  kickstart carries no farms and is separated from $T^{(5)}$ by mountains, so
  its farmland is inert and the playable area is precisely $T^{(5)}$. Taking
  the tile stock to be the exact multiset of non-starting tiles used, this
  order is legal under both editions: in the 2005 edition the tiles are drawn
  in this order, whilst in the 2022 edition the first four tiles fill the open
  display, the next required tile is chosen each turn, and the display is
  refilled in order until empty. Finally, the history has $O(|V(T)|^2)$
  placements and can clearly be constructed in polynomial time.
\end{proof}

\begin{proof}[Proof of \cref{thm:reachable-fjords}]
  Take the underlying tree of the reduction in \cref{subsec:gfjords-reduction},
  before the subdivision. Applying \cref{lem:honeycomb-drawing} draws its
  $r^2$-subdivision in the honeycomb grid for some polynomially bounded $r$, so
  each edge of the tree becomes a path of $r^2 + 1$ honeycomb edges. Our glove
  lemma (\cref{lem:glove}) replaces each honeycomb edge by a path of six edges
  (i.e.\ it takes a uniform 5-subdivision). Each edge of the underlying tree
  therefore becomes a path of $6(r^2 + 1)$ edges, so we choose $\ell = 6(r^2 +
  1) - 1$ in our reduction.

  The playable area is now precisely the tree for which we chose the initial
  colours. By \cref{lem:glove}, we can place a farm of the prescribed colour at
  each initially coloured vertex, and we can also arrange for Right to start
  the expansion phase. This gives a legal history of polynomial length for the
  exploration phase, which can be constructed in polynomial time using only
  four tile designs besides the mandated three-tile start. Note that both the
  playable area and its initially unoccupied subgraph are subcubic trees.

  Until a player first has no move, the expansion phase plays exactly as in
  normal play \GFjords. When a player first has no move, their opponent has
  placed at least as many fields. A player who cannot move will not be able to
  move for the remainder of the game,\footnote{
    In combinatorial game theory, this property is often called the
    \emph{dead-ending} property, and it is well-studied in partizan mis\`ere
    theory.
  } since they can only obtain new moves by playing moves themselves. Supplying
  each player with more fields than there are vacancies (with no fields placed
  initially), every vacancy in the playable area will eventually be filled by a
  field; if one were to remain, then a path from a farm to that vacancy must
  contain an occupied tile adjacent to an empty tile.

  By the rules of the game, only fields score, and the initial number of
  vacancies is odd. Thus, the normal play winner finishes with a strictly
  higher score under either published edition. The construction takes
  polynomial time (since $r$ and $\ell$ are polynomial in the input size), and
  the expansion phase requires only polynomial space since each move fills a
  tile.
\end{proof}

\section{Mis\`ere play}
\label{sec:misere}

Under mis\`ere play, the first player unable to move wins. As mentioned in the
introduction, we adapt an idea from Stewart's reduction for
\ruleset{Hackenbush}~\cite[Theorem~4]{stewart:misere-hackenbush}. In
particular, we add a \emph{poison} move that is constructed so that the first
player who makes it will lose the game. The careful check that we have to make
in our case is that the poison move is actually available to the player who
first runs out of moves elsewhere.

\begin{proof}[Proof of \cref{thm:graph-fjords} under mis\`ere play]
  Start with the same underlying tree as in the normal play construction,
  before the subdivision occurs, and write $N$ for its number of vertices. Now
  attach a new vertex $p$ (to be our \emph{poison vertex}) to the first vertex
  $z$ beyond $y$ on the reserve path. Give $p$ a private blue leaf and a
  private uncoloured path of $N+1$ edges.

  Now take the $\ell$-subdivision as before. In the new path between $p$ and
  what used to be its private blue leaf, make every internal vertex blue, so
  that $p$ still has a private blue neighbour. Count the $\ell$ new vertices
  between $z$ and $p$ as part of the reserve, shortening its uncoloured portion
  on the original path by the same amount. Since $M \ge s = (\ell+1)(4n)^2 >
  2(\ell+1)$ in the normal play construction, the shortened uncoloured portion
  of the original path still reaches past $z$ (and its red end only grows).
  Thus, $z$ remains uncoloured, and the reserve is still a tree of precisely
  $M$ uncoloured vertices meeting its red end; in particular, the normal play
  strategies and their move counts still hold until the moment that $p$ is
  taken. Both the entire graph and its initially uncoloured subgraph are still
  subcubic trees.

  Note that there are fewer than $N(\ell+1)$ uncoloured vertices outside $p$
  and its private path, whereas its private path alone has $(N+1)(\ell+1)$
  uncoloured vertices. Whoever claims $p$ becomes the only player able to move
  into its private path, securing more moves there than their opponent can make
  anywhere else. Hence, the opponent runs out of moves first and wins under
  mis\`ere play.

  It remains only to check who is forced to take $p$. If Right has the winning
  strategy in the normal play construction, then he follows it, avoiding $p$,
  until Left runs out of moves elsewhere; since $p$ has a private blue
  neighbour, Left is then forced to take $p$. If instead Left has the winning
  strategy in the normal play construction, then she follows it without
  entering Right's reserve or claiming $p$. Since the reserve is a connected
  subtree meeting its red end and Left never touches it, Right always has a
  legal move inside the reserve for as long as any of its vertices remains
  uncoloured. Hence, when Right has no legal move other than $p$, the entire
  reserve is already red; in particular, the neighbour of $p$ in the reserve is
  red, so Right must take $p$. In either case, the player who wins in the
  normal play construction also wins under mis\`ere play.
\end{proof}

\begin{corollary}
  \label{cor:tile-play}
  Under the same restrictions as \cref{thm:reachable-fjords}, deciding whether
  the second player wins the expansion phase is \PSPACE-complete under normal
  play, mis\`ere play, and mis\`ere scoring, even with only four tile designs
  besides the three mandated starting tiles.
\end{corollary}

\begin{proof}
  Normal play follows directly from the proof of \cref{thm:reachable-fjords},
  since expansion plays identically to normal play \GFjords until a player
  first has no move. For mis\`ere play, apply \cref{lem:honeycomb-drawing} to
  the enlarged underlying tree from the proof above (for some polynomially
  bounded $r$) and then \cref{lem:glove}, choosing $\ell = 6(r^2 + 1) - 1$ in
  the mis\`ere construction. The playable area, farms, and player to move of
  the resulting reachable board reproduce the mis\`ere \GFjords position
  exactly, so the same strategy argument applies.

  The same reachable board also yields hardness under mis\`ere scoring (where
  the strictly lower score wins). By the same argument as above, the normal
  play winner can avoid $p$ until their opponent is forced to claim it. Whoever
  claims $p$ becomes the only player able to move into its private path, and
  since each player has enough fields, they must eventually fill that entire
  path. Thus, they place strictly more fields than their opponent can possibly
  place elsewhere, so they finish with a strictly higher score (ruling out
  ties) and lose under mis\`ere scoring.
\end{proof}

Finally, the drawing from \cref{lem:honeycomb-drawing} also resolves the
question from the introduction on the honeycomb and triangular grids.

\begin{corollary}
  \label{cor:grid-fjords}
  Under both normal and mis\`ere play, \GFjords is \PSPACE-complete on
  subgraphs (indeed, subcubic trees) of the honeycomb grid and of the
  triangular grid.
\end{corollary}

\begin{proof}
  Applying \cref{lem:honeycomb-drawing} to the underlying tree of the normal or
  mis\`ere construction embeds its $r^2$-subdivision into $\infhive$ for some
  polynomially bounded $r$; choosing $\ell = r^2$ therefore realises the
  corresponding \GFjords instance as a subcubic tree in $\infhive$. Since
  $\infhive$ is an induced subgraph of $\inftri$, the same instance is also a
  subcubic tree in the triangular grid.
\end{proof}

\section{Subdivisions and forbidden graphs}
\label{sec:subdivisions}

We used long paths to draw the reduction using tiles. The freedom to choose
their length has a second use for us: for each fixed $\ell$, we can require
every edge of a subcubic tree to be subdivided exactly $\ell$ times and still
obtain hard instances.

\begin{theorem}
  \label{thm:subdivisions}
  If $\ell \ge 0$ is fixed, then both normal and mis\`ere play \GFjords are
  \PSPACE-complete on $\ell$-subdivisions of subcubic trees. The same holds for
  reachable \TFjords positions under the scoring rules of both published
  editions, under mis\`ere scoring, and under both normal and mis\`ere play,
  when the restriction is on the playable area.
\end{theorem}

\begin{proof}
  For \GFjords, use the given $\ell$ directly in the normal and mis\`ere
  constructions.

  For \TFjords, take the underlying tree of the corresponding construction and
  subdivide each edge $\ell$ times before drawing it. Applying
  \cref{lem:honeycomb-drawing} (for some polynomially bounded $r$) and then
  \cref{lem:glove} represents each edge of the original tree by a path of
  $6(r^2 + 1)(\ell + 1)$ edges. Apply the corresponding \GFjords construction
  with subdivision depth $6(r^2 + 1)(\ell + 1) - 1$. The resulting playable
  area is itself an $\ell$-subdivision of a subcubic tree (splitting each such
  path into $6(r^2 + 1)$ paths of length $\ell + 1$), and reachability and the
  winning strategies follow just as before.
\end{proof}

For the graph game \GFjords, we may even prescribe the subdivision depth as a
fixed function of the order of the underlying tree of the construction,
provided that this function is polynomially bounded and computable in
polynomial time.

Johnson et al.\ have studied how the complexity of a graph problem changes when
a finite family of subgraphs is
forbidden~\cite[Section~1.1]{johnson.ea:forbidden-subgraphs}. In particular,
they develop what they call the C123 framework. The first condition (C1)
requires tractability on graphs of bounded treewidth, the second (C2) asks for
hardness on subcubic graphs, and the third (C3) asks for hardness to persist at
arbitrarily large fixed uniform subdivision depths.

Thus, both normal and mis\`ere \GFjords satisfy C2 and C3. Since hardness
already holds on trees, which have treewidth~$1$, we know that C1 fails unless
$\P = \PSPACE$. The same holds for reachable \TFjords positions under all four
conventions above (scoring under either published edition, mis\`ere scoring,
normal play, and mis\`ere play), where the restriction is on the playable area
of the position. In this way, we obtain several new examples of C23 problems.

Let $\set{S}$ be the set of graphs whose connected components are paths
(including single vertices) or subdivided claws. As in the forbidden-subgraph
framework~\cite{johnson.ea:forbidden-subgraphs}, C2 and C3 yield the following
consequence.

\begin{corollary}
  \label{cor:forbidden-subgraphs}
  If $\set{H}$ is a finite family of graphs with
  $\set{H} \cap \set{S} = \emptyset$, then \GFjords (under normal and mis\`ere
  play) and reachable \TFjords (under standard scoring in either published
  edition, mis\`ere scoring, normal play, and mis\`ere play) are
  \PSPACE-complete on $\set{H}$-subgraph-free graphs (and also $\set{H}$-free
  graphs), even restricted to subcubic trees (with this restriction placed on
  the playable area for \TFjords).
\end{corollary}

\begin{proof}
  Apply \cref{thm:subdivisions} with any integer $\ell \ge 0$ at least the
  order of every graph in $\set{H}$. Every graph outside $\set{S}$ contains a
  cycle, a vertex of degree at least four, or two vertices of degree three in
  the same component. The resulting $\ell$-subdivisions of subcubic trees are
  acyclic, have maximum degree at most three, and have any two vertices of
  degree three at distance at least $\ell + 1$ (exceeding the order of every
  graph in $\set{H}$). Hence, no graph in $\set{H}$ occurs as a subgraph, nor
  in particular as an induced subgraph.
\end{proof}

For minors and topological minors, there is a further consequence. Every minor
of a tree is a forest, and every topological minor of a subcubic tree is a
subcubic forest. Hence, our hard instances exclude $H$ as a minor whenever $H$
contains a cycle, and as a topological minor whenever $H$ contains a cycle or a
vertex of degree at least four. For ordinary subgraph exclusion, the graphs $H$
outside the scope of \cref{cor:forbidden-subgraphs} are precisely those whose
components are paths or subdivided claws.

It remains open to classify the complexity of \GFjords and reachable \TFjords
(under any of our scoring or play conventions) for the remaining forbidden
graphs $H$, both for $H$-subgraph-free graphs and for the other standard graph
containment relations ($H$-minor-free, $H$-topological-minor-free, $H$-free,
$H$-induced-minor-free, and $H$-induced-topological-minor-free graphs). Another
open question concerns the number of farms: our exploration histories for
\TFjords use a growing number of farms, and we do not know whether hardness
persists when the number of farms is bounded (even if the supply of tiles and
fields is still allowed to grow).

\pdfbookmark[1]{References}{references}
\bibliography{ref}

\end{document}